# Calibration of off-the-shelf low-cost wearable EEG headset for application in field studies

Manvi Jain[1]*, C.M. Markan[2]

[1] Department of Cognitive Sciences, Dayalbagh Educational Institute, Agra, 282004, India
[2] Department of Physics and Computer Science, Dayalbagh Educational Institute, Agra 282004, India

*Electroencephalography (EEG) is an integral tool in neurocognitive research worldwide. However, research grade EEG (32/64ch) systems are expensive and have cumbersome setup designed for clinical usage not suited for rugged environment of field-studies outside lab. Further, the long setup-time of EEG can be intimidating to those who are restless subjects e.g., children or elderly. Off-the-shelf, low-cost, dry EEG devices (LCDE) have been proposed as promising options. However, small number of electrodes in LCDE limit the detection scalp-area reducing the utility of an LCDE only to a specific set of cognitive tasks based on the brain lobe scanned. This paper proposes a novel methodology for calibration of an LCDE (e.g., DREEM Headband) to identify the specific class of cognitive tasks a LCDE is likely suited for. The methodology involves comparative analysis of the recorded data using LCDE with EEG-like signals simulated (using BESA Simulator software) by embedding dipole in a brain lobe. The simulated scalp activity and corresponding source analysis helps identify the approximate regions of brain scanned by the LCDE device. On further comparative analysis of brain lobes source localized using Brain Electrical Source Analysis (BESA software) helps characterize LCDE for cognitive tasks. The major findings conclude a list of psychological studies which can be performed using various LCDE, capable of replacing traditional and expensive wet EEG systems in both inside and outside-lab settings.*

**Keywords** – *Dry EEG, Dreem headband, localization error, BESA, dipole simulation, psychological testing*

## I. INTRODUCTION

Electroencephalography (EEG) signals reflect physiological markers of cognition and awareness (LaRocco et al., 2020). However, traditional EEG systems extensively used in research today are usually very expensive and sensitive, more importantly, depend upon the use of dozens of channels, and have detailed setup time, rendering their use impractical for implication in outside-lab settings (Puce & Hämäläinen, 2017). In such condition, LCDE shown in Fig. 1 could be recruited in neuropsychological studies for obtaining results similar to that recorded using standard wet EEG systems (Arnal et al., 2020; LaRocco et al., 2020; Puce & Hämäläinen, 2017).

The devices shown in Fig. 1 are advantageous over traditional EEG systems due to several reasons: 1) these are headband-like devices which can be carried easily in the field settings, 2) plugin-play i.e. easy, faster and user-friendly set-up system, 3) lesser number of channels to focus on specific brain regions depending upon the task displayed, and 4) can be used in studies performed on young children when cumbersome set-up of traditional EEG is difficult (Johnstone et al., 2020). The LCDE to be used must be chosen on the basis of: availability and accessibility of resources, set-up time required, and minimum technical experience required to use the device.

However, research shows that there is a limitation to cognitive activities that can be accounted for with such devices due to the limited number of channels (Puce & Hämäläinen, 2017; Radüntz, 2018). Most of all the psychological tests performed using EEG devices involve certain cognitive functioning deriving from specific brain regions. The fact that LCDE contain a smaller number of channels becomes important because usually psychologists find it difficult to identify the electrodes specific to the region of interest in the brain according to their test, ending up extracting a data full of noise from other unimportant channels in the EEG system. To identify a solution to this problem, section III of this paper has studied simulated activation produced in different brain lobes to recognize the channels that can record it. Result of section III is a chart of cognitive studies which can be performed using LCDE available based on the cognitive activity localized in specific brain lobes for future reference while identifying a suitable dry EEG device corresponding to the psychological tests applied. Example, visual and basic cognitive tasks based on variables like reaction time and basic analytical problem-solving skills are the most suitable for analysis using devices that cover frontal brain region such as DREEM headband and Emotiv epoc+ (Fig. 1). Some low-cost and easily accessible LCDE are reviewed briefly on the basis of the number of channels and major characteristic features below:

**DREEM Headband:** The headband-like dry EEG device covers frontal and occipital regions of brain with electrodes placed at F7, F8, FP1, O1, O2 and Fp2 as ground. Signals are sampled at 250 Hz. Along with electrodes, it also consists of a pulse sensor to monitor

---

*Corresponding author: manvijain65@gmail.com

heart rate and accelerometer to measure head movements and respiratory rate [3].

**Emotiv epoc+:** The device is extensively used in research in the fields of Brain-computer interface and brain state detection (Badcock et al., 2013). It consists of two electrode arms, each comprised of several sensor electrodes along with two reference electrodes covering the frontal, temporal, parietal, and occipital lobes of brain at following electrode positions- Frontal (Fp1, Fp2, F3, F4, F7, F8, FC5, FC6); Temporal (T7, T8); Parietal (P3, P4, P7, P8) and Occipital (O1, O2).

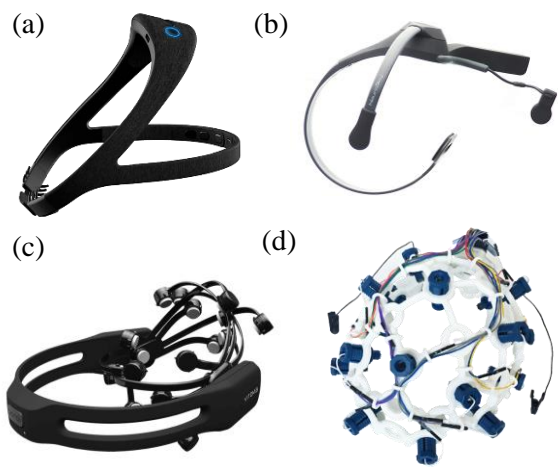

Fig. 1: Low-cost Dry EEG (LCDE) devices: a) DREEM headband [16], b) Neurosky mindwave mobile [17], c) Emotiv epoc+ [18] and d) OpenBCI Ultracortex Mark IV [19]

**Neurosky Mindwave mobile:** The single-channel wireless device is a low-cost, dry EEG headset, capable of transmitting EEG signals from electrodes of interest via Bluetooth to smartphone (Doudou & Bouabdallah, 2018). The headset consists of a single electrode over Frontal lobe (Fp1) on forehead above left eye.

**OpenBCI Ultracortex Mark IV:** The device samples up to 16 EEG channels using dry EEG sensors with set-up time almost similar as a wet EEG system It is capable of recording brain electrical activity with EEG, muscle activity with EMG (electromyograph) and heart activity with ECG (electrocardiograph) (Mohamed et al., 2018).

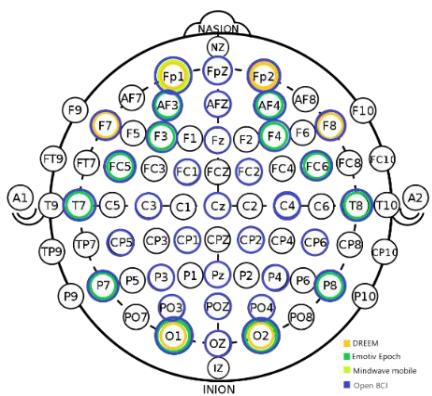

Fig. 2: Electrode map for the four dry EEG devices (LCDE) *(Electrode Map, n.d.)*

### A. Motivation of study

Lesser number of channels in dry-EEG devices restrict data acquisition to specific brain regions. According to Michel and colleagues (2004), change in brain activity at electrodes of interest produces artefactual activity at all electrodes irrespective of their position. However, activity at all electrodes irrespective of their position. However, activity at all electrodes cannot be recorded by single or few-channeled devices, preventing analysis of major changes in brain regions proximal to regions of interest. This demands for a possible method for characterization of such devices before their implementation in practical world.

This describes the aim of the study to characterize the quality of data collected by one such device- DREEM headband in comparison to near to ideal data, simulated using dipole simulator program of brain electrical source analysis (BESA) software.

### B. Contributions

The main contributions of this paper are as follows:
- To simulate cortical activity in the four brain lobes shown in Fig. 3 (a) using BESA simulator by embedding dipoles in a few electrode positions to localize the corresponding source of activity in the brain. This is required for illustrating the combinations of minimum electrodes that can be used for studying the neural activity evoked by the psychological test-specific cognitive functions and organize a list of tests which can use devices containing channels in region of interest.
- To devise a methodology for comparison between EEG signals collected by the DREEM headband and standard noise-free signals simulated using BESA simulator at frontal and occipital lobes of the brain, for calibration of data quality derived from DREEM headband in comparison to the simulated dataset based upon source localization method.

## II. MATERIALS

### A. Study Resources-

**DREEM headband ©DREEM 2021:** The DREEM device is a wireless headband-like EEG device that records electrical signals produced in brain during different stages of sleep without requiring any external connections. It records brain cortical activity with 5 (+1 ground) dry electrodes and provide 7 derivations (FpZ-O1, FpZ-O2, FpZ-F7, F8-F7, F7-O1, F8-O2, FpZ-F8); the sampling frequency is 250 Hz; and the signals are bandpass filtered at 1Hz–30 Hz.

**Brain electrical source analysis (BESA)** ("BESA,"

n.d.)**:** Dipole simulator program produced by Patrick Berg allows simulation of evoked or induced brain activity (Berg & Scherg, 1994). BESA dipole simulator allows dipole approximation or modelling (Litt, 1991; Miltner et al., 1994) which simulates electrical activity similar to that generated by partial depolarization of pyramidal neurons in the brain. Dipoles in the brain as shown in Fig. 3 (b) are caused by charge separation in neurons during flow of electrical charges for information signaling (Salvetti & Wilamowski, 2008).

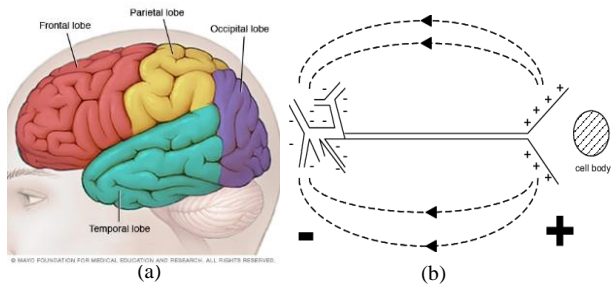

Fig. 3. a) The four lobes of brain: Frontal, Parietal, temporal and Occipital. *Source: Mayo Foundation.* b) A dipole model generated at the apical dendrite tree

### III. ILLUSTRATIVE STUDY: SIMULATION

#### A. Simulation Procedure:

Fig. 2 is an electrode map representing all electrode positions covered by the four LCDE discussed in Section I. We selected only few major electrodes positions out of all available data for all four lobes and simulated activity using minimum two dipoles and maximum six dipoles to describe the corresponding source localization in each brain lobe.

Positions of dipoles placed in the brain lobes during simulation process are: frontal lobe, six: Fp1, Fp2, Fpz, F3, F4, F7, F8; parietal lobe: Pz, P3, P4; temporal lobe: T5, T6 and occipital lobe: O1, O2.

#### B. Simulation results:

The simulated waveforms for dipoles embedded in four brain lobes (resulting figures given in supplementary material I) were imported in BESA software for spatial analysis using 1) current source density maps which depict surface activity averaged over short segments of time, shown in Fig. 4. (c), and 2) distributed imaging method called Classical LORETA Analysis Recursively Applied (CLARA) which is an iterative application of famous sLORETA technique of source analysis [14].

Resulting surface activity for frontal lobe simulation is represented as a colormap depicting regular activity in and near frontal regions of the head model, depicted in Fig. 4. (c) showing six surface activity maps for 30 seconds time duration. The source analysis for the frontal lobe simulation is represented in Fig. 4. (b) in which source activity specific heightened power at location of simulation i.e., frontal region of the brain can be clearly seen.

This shows that similar simulation procedure can be repeated for any dry EEG device to estimate the region of resulting activation to compare whether the device is able to identify the activation in the region of interest. Conclusively, we have created a list of possible psychological tests that could be performed using the available LCDE(s) after applying similar simulations in some regions. We have generalized the solution by using source localization technique as most researchers prefer deeper cortical analysis due its time independent nature. Table I (at the bottom) is a list of psychological tests composing the related cognitive functioning and region of activation in brain (as suggested by past studies). Through the analysis, we have suggested minimum number of electrodes that can identify neural activation in the tests along with a suitable dry EEG device.

Another application of such simulation techniques is in studies which aim to create a customizable BCI design for advanced studies such as those which target at a specific brain region to accumulate brain electrical activity in order to transform the signals into a visible form such as converting neural signals into handwriting or other signal to image processing methods.

### IV. COMPARITIVE STUDY: METHODOLOGY

#### A. Data collection & pre-processing

Task-related brain activity was recorded using DREEM headband device on a human subject (with no recent neuropsychological or visual disorder history) for 4 minutes while the subject was involved in a basic visual-cognitive task called Tower of London task which induces activity located in frontal region (as it requires progressive planning and decision making) and occipital region (since it is a visual task, it requires active visual perception) of the brain.

The distribution of six electrodes in the device follows the 10-20 international system of electrode placement: channels Fp1, Fp2 (reference channel), F7, F8 in frontal region and O1, O2 in occipital region are present. The recorded dataset was imported in BESA for artefact identification through its automatic Pattern search tool for identifying regular eye blinks and cardiac activity and removing them for artefact reduction. For noise filtering, the dataset was band passed at 0.1Hz-30Hz for specifying frequency content.

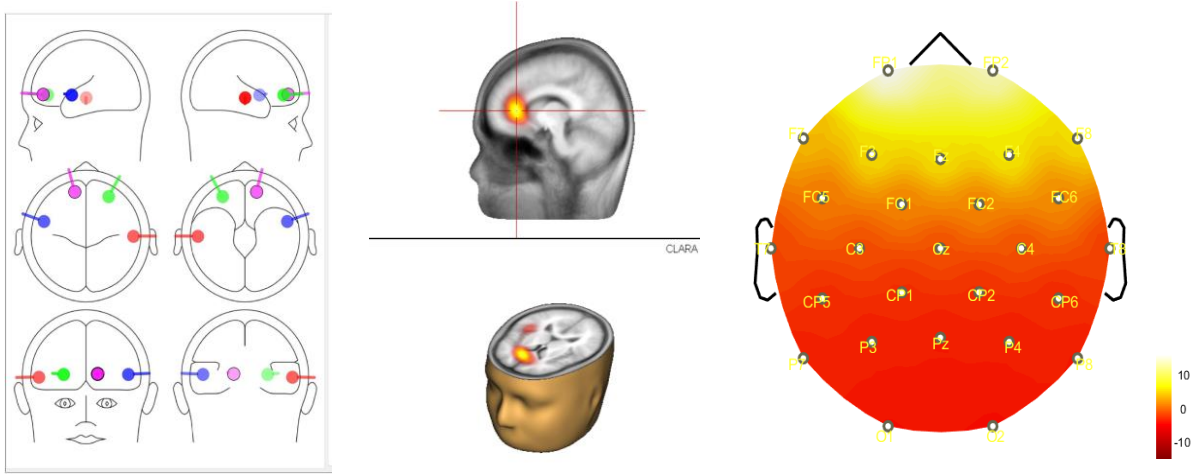

Fig. 4 a: Frontal lobe: (a) dipoles embedded in frontal region of head model. (b) Source localization depicting activation in frontal lobe. (c) Cortical or surface activity colormap for dataset simulated for dipoles placed at frontal lobe depicting heightened surface activity over frontal region of brain.

### B. Simulation of model dataset-

The previous section (Section III) illustrated the inverse mapping feature of artificial dipole placement using BESA dipole simulator. Six artificial dipoles were seeded in a head model at electrode positions same as that of DH to simulate a) noise-free b) rms noise voltage: 0.2µV c) rms noise voltage: 0.4µV, d) rms noise voltage: 0.6µV, e) rms noise voltage: 0.8µ datasets. The embedded dipoles and their corresponding waveforms with different noise levels can be found in supplementary material II. For application of this tool, we identified the MNI coordinates of all six electrode positions exhibited by DREEM Headband for simulation in a standard head model in frontal and occipital region of the model.

### C. Noise sensitivity of model dataset(s):

Noise sensitivity for EEG-like signals can be determined by several methods. The mathematical concept used in this study to determine noise sensitivity is derived using localization error resulting from source analysis of model dataset simulated at different noise levels. Here, the localization error states the difference between the localized source for noise-free dataset and noisy dataset (at different noise levels), both simulated using BESA dipole simulator program. Steps followed for measuring noise sensitivity level of model dataset(s) are: 1) Source analysis using CLARA technique of localization was performed in BESA software, resulting localization figures can be found in Supplementary material III  2) the resulting MNI coordinates for highest amplitude source localized by the software were tabulated. 3) Localization error ≪ε≫ was calculated based on the equation used in previous studies (Khemakhem et al., 2017) in MATLAB 2020a. The resulting noise sensitivity for various model dataset(s) is represented in Fig. 5.

$$\varepsilon = \sqrt{(x^1 - x^2)^2 + (y^1 - y^2)^2 + (z^1 - z^2)^2} \quad (1)$$

The resulting graph in Fig. 5. shows that there is a trend in increase in noise sensitivity with increasing noise level in model dataset, starting with minimal noise sensitivity in model dataset (b) and (c) as 0.122 and 0.126 respectively, drastically increasing to 1.077 and 1.071 in model dataset (d) and (e).

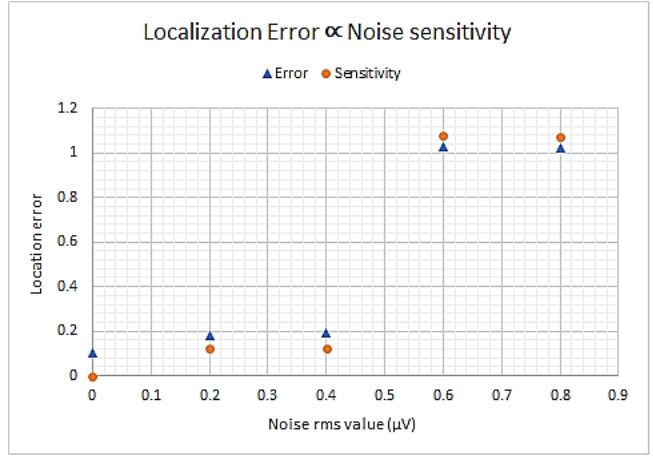

Fig 5. Noise sensitivity (represented by orange circles) for model datasets (b-e) with respect to dataset (a) and localization error (represented by blue triangles) for model datasets with different noise levels with respect to dry EEG data.

### V. RESULTS

#### A. Comparative analysis between DREEM Data and model dataset

For comparison with dry EEG data recorded using DREEM Headband, we selected the noise-free model dataset amongst all the datasets simulated using BESA dipole simulator program. Both the comparison datasets were free of artefacts and bandpass filtered for normalization purpose. The method for comparison included source analysis in BESA software to compare the localized source in both datasets and determine the level of localization error to establish a correlation between both the datasets. The source analysis technique used was CLARA which is Classical LORETA Analysis Recursively Applied since localization error is minimum in sLORETA amongst all source analysis techniques (Pascual-Marqui, 2002). The localized source for both comparable datasets can be seen in Fig. 6. and the localization error between the two sources, calculated using equation 1 comes out to be 0.1871 depicting minimal level of deviation between the two localized sources.

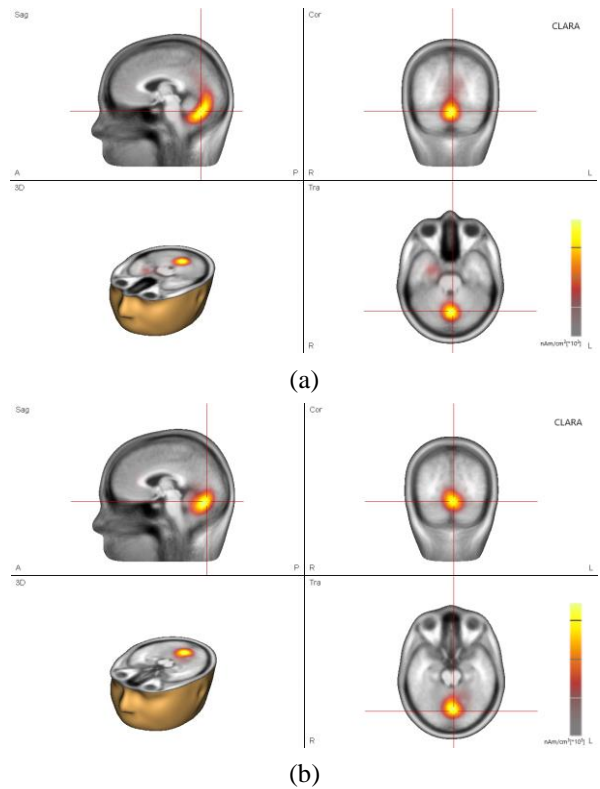

(a)

(b)

Fig. 6: Source localization for comparative analysis between (a) Noise-free model dataset simulated in BESA dipole simulator program and (b) Dry EEG dataset recorded using DREEM Headband

## VI. CONCLUSION

The results of the comparative study shows that a dry EEG data in its processed state (i.e., computational artefactual corrections applied), is comparable with an artificially simulated noise-free data with a minimal localization error of 10%. This depicts that the dry EEG data is free of any sort of environmental, muscle movement or any other noise created by the surrounding equipment. Field Studies performed outside sophisticated lab environment require the use of less sensitive, portable and low-cost neurophysiological monitoring equipment. However, the data quality might be affected because of ruthless environment of monitoring, so the proposed method of comparative analysis in this study could be a good calibration methodology for characterizing any brain activity signals recorded using LCDE before their real-world implementation to account for the quality of data recorded using the devices.

A more important conclusion of this study is a reference chart of psychological tests suggesting the EEG electrodes that can be enforced in further cognitive studies that aim to identify source of activation in the brain resulting pertaining to the cognitive function involved in the tests, along with a list of preferred dry EEG device to use.

## VII. DISCUSSIONS

The analytical results of this study provide a scope for comparing signals derived using a dry EEG device – DREEM Headband with a model data simulated using BESA software. The scope of application of easily available and portable LCDE such as DREEM Headband is proposed in a study design based upon identifying neurophysiological activity of infants or children in early childhood to keep a track of their developmental patterns. More elaboratively, our lab in Dayalbagh Educational Institute is preparing to work on a related proposal with children below three-five years of age in fields of Dayalbagh while performing different cognitive tasks. Before real-world implementation of the LCDE on such young children, it became possible to calibrate the data collected by the device to understand and rectify the issues involved in advance.

Limitations of the study: Future studies could use wet EEG data for comparison with Dry EEG data and model dataset for finding the correlation between datasets recorded using wet EEG, dry EEG and model data simulated using BESA simulator.

TABLE I

DRY EEG DEVICES FOR COGNITIVE TESTING

| Cognitive Tasks | Lobes of Activation | Cognitive functions | Appropriate Electrodes | Devices |
|---|---|---|---|---|
| Stroop task | Right prefrontal cortex (Vendrell et al., 1995) | Executive functions; visuospatial processing, colour determination, sustained attention | Frontal: Fp1, Fpz, F7, F8, F3, F4, FC5, FC6; Occipital: O1, O2 | DREEM Headband, Emotiv epoc+ |
| Number-letter task | Orbital frontal lobe, dorsolateral prefrontal cortex, and posterior parietal cortex (Haut et al., 2000) | Visualization of the verbal information, working memory | Frontal: Fp1, Fp2, AF3, AF4, Fpz, AFz, Fz, F7, F8, C3; Parietal: Cz, C4, CP5, CP1, CP2, CP6, Pz, P7, P8, P3, P4 | Open BCI |
| Letter-memory task | Lateral prefrontal cortex (Owen et al., 1998) | Spatial and nonspatial working memory, visualization | Frontal: Fp1, Fp2, F3, F4, F7, F8, FC5, FC6; Occipital: O1, O2 | Emotiv epoc+ |
| Anti-saccade task | Frontal lobe (dlPFC-dorsolateral prefrontal cortex) (Klein et al., 2010) | Working memory, executive functions, and intelligence | Frontal: F3, F4, Fp1, Fp2, F3, F4, F7, F8, FC5, FC6 [Normal Electrical Activity of the Brain in Obsessive-Compulsive Patients After Anodal Stimulation of the Left] | Emotiv epoc+ |
| Wisconsin card sorting task | Frontoparietal (Buchsbaum et al., 2005) | Set shifting, task switching, working memory and inhibitory control | Frontal: Fp1, Fp2, F3, F4, F7, F8; Parietal: P3, P4, P7, P8 | Emotiv epoc+ |
| Tower of hanoi | Frontal | Goal-subgoal conflict resolution, working memory (Goel & Grafman, 1995) | Frontal: Fp1, Fpz, F7, F8, **F3, F4, FC5, FC6**; Occipital: O1, O2 | DREEM Headband, Emotiv epoc+ |
| Hooper visual organization test (HVOT) | Frontal (precentral gyrus), parietal and occipital (Jefferson et al., 2006) | Executive functioning, lexical retrieval, visuospatial processing | Frontal: Fp1, Fpz, F7, F8, **F3, F4, FC5, FC6**; Parietal: Cz, C4, CP5, CP1, CP2, CP6, Pz, P7, P8, P3, P4; Occipital: O1, O2, PO3, PO4, POz, | Open BCI |
| Clock drawing task – clox (1 and 2) | Bilateral frontal, occipital and parietal lobes (Talwar et al., 2019) | Visuospatial processing, executive function, semantic memory, and planning | Frontal: Fp1, Fpz, F7, F8, **F3, F4, FC5, FC6**; Parietal: Cz, C4, CP5, CP1, CP2, CP6, Pz, P7, P8, P3, P4; Occipital: O1, O2, PO3, PO4, POz, | Open BCI, Emotiv epoc+ |
| D2 test | Frontal (Brickenkamp, 1998) | Individual attention and concentration performance | Frontal: Fp1, Fp2, F7, F8; Occipital: O1, O2 | Neurosky mindwave mobile, DREEM Headband |
| Target word search (WS task) | Frontal, occipital (Fockert et al., 2004) | Visual processing, logical reasoning, problem-solving | Frontal: Fp1, Fp2, F3, F4, F7, F8, FC5, FC6; Parietal: P3, P4, P7, P8; Occipital: O1, O2 | Emotiv epoc+ |
| Trail-making test A&B | Frontal cortex (Müller et al., 2014) | Attention-related functions | Frontal: Fp1, Fp2, F7, F8; Occipital: O1, O2 | Neurosky mindwave mobile, DREEM Headband |
| Visual and auditory continuous performance test (IVA) | Frontal, parietal and occipital | Vigilance, auditory focus, auditory speed, auditory consistency, visual focus, visual Speed, visual prudence, and visual consistency (Tinius, 2003) | Frontal: Fp1, Fpz, F7, F8, **F3, F4, FC5, FC6**; Parietal: Cz, C4, CP5, CP1, CP2, CP6, Pz, P7, P8, P3, P4; Occipital: O1, O2, PO3, PO4, POz, | Open BCI |

Table footnote: List of psychological tests along with the cognitive functions and region of activation in brain corresponding to them (as suggested by past studies). Using results of source analysis, minimum number of electrodes that can identify neural activation in the tests are suggested along with a suitable LCDE which is both easily accessible and can provide quality results.